# Study the function of building blocks in SHA Family

*A. Arul Lawernce selvakumar[1], Dr.R.S.Ratastogi[2], Member IAENG*

*Abstract:* **In this paper we analyse the role of some of the building blocks in SHA-256. We show that the disturbance correction strategy is applicable to the SHA-256 architecture and we prove that functions Σ, σ are vital for the security of SHA-256 by showing that for a variant without them it is possible to find collisions with complexity $2^{64}$ hash operations. As a step towards an analysis of the full function, we present the results of our experiments on Hamming weights of expanded messages for different variants of the message expansion and show that there exist low-weight expanded messages for XOR-linearised variants.**

## I. INTRODUCTION

Recent results on the practical cryptanalysis of many hash functions from the MD family, including MD4, MD5 [WLF[+]05,WY05] as well as SHA-0 and SHA-1 [BCJ[+]05,RO05, WYY05b, WYY05a], drew a considerable attention to the security of hash functions and raised some questions about the security of the latest function in this family, namely SHA-256. The first published independent analysis of the members of the SHA-2 family was done by Gilbert and Handschuh Showed that there exists a 9-step local collision with probability $2^{-66}$. Later on, this result has been improved by Hawkes, Paddon and Rose [HPR04]. They showed how to increase the probability to $2^{-39}$ using modular differences.

In this paper we investigate the limits of applying the disturbance-correction strategy that was introduced by Chabaud and Joux [CJ98] to cryptanalyse SHA-0. We demonstrate the importance of the S-boxes applied in SHA-256. Throughout this paper. we use different linearisation models, namely a linearisation with respect to the XOR-operation (XOR-linear) and a linearisation with respect to modular addition (ADD-linear). We start from the analysis of an ADD-linear variant of SHA-256 and derive a differential characteristic that produces collisions for that linear model. Next, we present a zero-output differential characteristic with probability $2^{-64}$ for the hash function with the Boolean functions. This proves that the application of the functions $\Sigma_0, \Sigma_1, \sigma_0$, and $\sigma_1$ is crucial for the security of the original hash function, since they are replaced by the identity function in this analysis. In parallel to this work, a different variant of SHA-256 was analysed by Yoshida and Biryukov [YB05].

A better understanding of the impact of these functions on the whole design is the next step in the analysis of SHA-256. While the influence of $\Sigma_0$ and $\Sigma_1$ on the probability of single correction has been studied well by Hawkes et al. [HPR04], as far as [GH03]

Manuscript received January 5, 2009. This work is supported for my Ph.D., Research work and supported in Department of Computing Engineering, Magadh University, INDIA.

They know, there has been no analysis of the message expansion involving $\sigma_0$ and $\sigma_1$. In this paper we discuss some properties of the message expansion and present our results of the search for low-weight message differences for various (XOR-linear) variants of the message expansion.

## II. DESCRIPTION OF SHA-256

SHA-256 [Nat02] is an iterated cryptographic hash function based on a compression function that updates the state of eight 32-bit chaining variables A,...,H according to the values of 16 32-bit words $M_0,..., M_{15}$ of the message. The compression function consists of 64 identical steps presented in Figure 1. The step transformation employs bitwise Boolean functions

$$\text{Maj}(A,B,C) = (A \wedge B) \vee (A \wedge C) \vee (B \wedge C),$$
$$\text{Ch}(E,F,G) = (E \wedge F) \vee (\neg E \wedge G),$$

and two S-boxes

$$\Sigma_0(x) = \text{ROT R}^2(x) \oplus \text{ROT R}^{13}(x) \oplus \text{ROT R}^{22}(x),$$

$$\Sigma_1(x) = \text{ROT R}^6(x) \oplus \text{ROT R}^{11}(x) \oplus \text{ROT R}^{25}(x),$$

built from word rotations to the right (ROT R) and bitwise XORs denoted by. The i-th step uses a fixed constant $K_i$ and the i-th word $W_i$ of the expanded message.

The message expansion works as follows. An input message is split into 512-bit message blocks (after padding). A single message block will be denoted either as a row vector $m \in Z_2$ or as a vector $M_t \in Z_2 32$ with $0 \leq i < 16$.

During the message expansion, this input is expanded into a vector of 64 32-bit words which may be also seen as the 2048-bit expanded message row-vector $w_i \in Z_2^{512}$. The words $W_i$ are generated from the initial message M according to the following formula:

$$W_i = \begin{cases} M_i & \text{for } 0 \leq i < 16 \\ \sigma_1(W_{i-2}) + W_{i-7} + \sigma_0(W_{i-15}) + W_{i-16} & \text{for } 16 \leq i < N \end{cases}$$

<div style="text-align:right">1</div>

*Figure1: One step of the SHA-256 compression function*

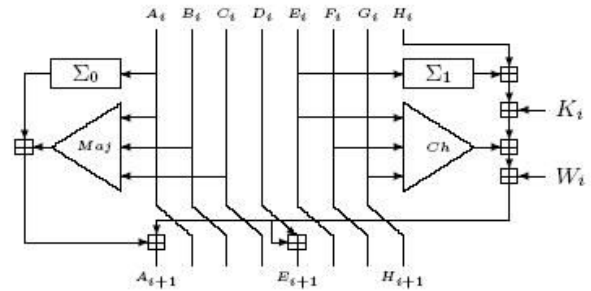

If we set N=64, we get standard SHA-256, taking a different value of N results in a reduced (or extended) variant of it. The functions $\sigma_0(x) = ROTR^7(x) \oplus ROTR^{18}(x) \oplus SHR^3(x)$ and $\sigma_1(x) = ROTR^{17}(x) \oplus ROTR^{19}(x) \oplus SHR^{10}(x)$ are S-boxes defined using word rotations to the right (ROT R) and shifts to the right (SHR).

## III. COMPUTING COLLISIONS FOR AN ADD-LINEAR VARIANT OF SHA-256

In order to analyse the usefulness of a disturbance-correction strategy applied to the SHA-2 architecture, we investigate an DD-linear variant of SHA-256, where S-boxes are replaced with the identity function, act which will be used later on.

$$\sigma_0 = \sigma_1 = \Sigma_0 = \Sigma_1 = id, \qquad (2)$$

And Boolean functions are replaced by the addition modulo $2^{32}$,

$$Maj(x, y, z) = Ch(x, y, z) = x + y + z. \qquad (3)$$

Now the whole function consists only of linear operations with respect to the modular addition. If we introduce a difference $\Delta_i = W_i^0 - W_i$, we can cancel this disturbance by introducing in the next 8 steps $i + 1, ..., i + 8$ the following sequence of corrections

$$\{-4\Delta_i, 2\Delta_i, 2\Delta_i, 4\Delta_i, 2\Delta_i, \Delta_i, 0, -\Delta_i\} \qquad (4)$$

The whole process of correcting a single disturbance is presented in Table 1. In the first 4 steps we use corrections that keep differences from influencing register A and later from step $i + 4$ we successively cancel differences in the register H.

The next step is to find a disturbance pattern $\Delta$ that follows the expansion process and can give raise to a corrective pattern. We will use an argument similar to the one used for finding disturbance patterns for SHA-1 [MP05, RO05, PRR05]. Let us introduce the necessary notation first. For any vector $s = [s_0, ..., s_l]$, let us denote by $\text{Delay}^a(s)$ a

| step s | $\Delta A_s$ | $\Delta B_s$ | $\Delta C_s$ | $\Delta D_s$ | $\Delta E_s$ | $\Delta F_s$ | $\Delta G_s$ | $\Delta H_s$ | $\Delta W_s$ |
|---|---|---|---|---|---|---|---|---|---|
| $i$ | 0 | 0 | 0 | 0 | 0 | 0 | 0 | 0 | $\Delta_i$ |
| $i+1$ | $\Delta_i$ | 0 | 0 | 0 | $\Delta_i$ | 0 | 0 | 0 | $-4\Delta_i$ |
| $i+2$ | 0 | $\Delta_i$ | 0 | 0 | $-2\Delta_i$ | $\Delta_i$ | 0 | 0 | $2\Delta_i$ |
| $i+3$ | 0 | 0 | $\Delta_i$ | 0 | $-\Delta_i$ | $-2\Delta_i$ | $\Delta_i$ | 0 | $2\Delta_i$ |
| $i+4$ | 0 | 0 | 0 | $\Delta_i$ | $-\Delta_i$ | $-\Delta_i$ | $-2\Delta_i$ | $\Delta_i$ | $4\Delta_i$ |
| $i+5$ | 0 | 0 | 0 | 0 | $\Delta_i$ | $-\Delta_i$ | $-\Delta_i$ | $-2\Delta_i$ | $2\Delta_i$ |
| $i+6$ | 0 | 0 | 0 | 0 | 0 | $\Delta_i$ | $-\Delta_i$ | $-\Delta_i$ | $\Delta_i$ |
| $i+7$ | 0 | 0 | 0 | 0 | 0 | 0 | $\Delta_i$ | $-\Delta_i$ | 0 |
| $i+8$ | 0 | 0 | 0 | 0 | 0 | 0 | 0 | $\Delta_i$ | $-\Delta_i$ |
| $i+9$ | 0 | 0 | 0 | 0 | 0 | 0 | 0 | 0 | |

*Table 1: Correcting a single disturbance $\Delta I$ introduced in step $i$ in an ADD-linearised variant of SHA-256*

Vector constructed by proceeding elements of s by a zero elements i.e.,

$$\text{Delay}^a(s) = [\underbrace{0, \ldots, 0}_{a\text{ times}}, s_0, \ldots, s_l]$$

And by delay $^a n(s)$ the same vector truncated to only n first elements i.e.,

$$\text{Delay}_n^a(s) = [\underbrace{0, \ldots, 0}_{a\text{ times}}, s_0, \ldots, s_{n-1-a}].$$

### LEMMA III.A

*Let $W \in Z_2 32$. If $\text{Delay}^a(W)$ is a result of the expansion using the recursive formula (1) with $N = 64 + a$, then all the vectors $\text{Delay}^b 64(W)$ for $0 \leq b \leq a$ are also results of the expansion process (1).*

PROOF: Each vector $\text{Delay}^b 64(W)$ consists of elements of the part of a sequence following the recurrence relation, also follows the relation.

Where $I_{16}$ stands for the identity matrix and A denotes a matrix of the linear transformation producing 16 new words out of 16 old ones according to the recurrence relation (1). The following theorem fully characterises disturbance patterns for an ADD-linear variant of SHA-256.

### THEOREM III.B

*Let $\Delta_M = M^0 - M$ be a message difference. The expanded difference $\Delta = E(\Delta_M)$ is a valid disturbance vector for an ADD–linear variant of SHA-256 if the following conditions are satisfied:*

$$0 = A^3[8::16] \cdot \Delta_M, \qquad (6)$$
$$0 = A_{-1}[8::16] \cdot \Delta_M, \qquad (7)$$

*where $M[a::b]$ means a matrix consisting of rows of the matrix M from the a-th row to the b-th row inclusive.*

PROOF. The fundamental observation is that each single word $\Delta_{ii}$ of the disturbance vector has to be corrected by adding to the next 8 words the following differences defined by Equation (4),

$$\{-4\Delta_i, 2\Delta_i, 2\Delta_i, 4\Delta_i, 2\Delta_i, \Delta_i, 0, -\Delta_i\}$$

This shows that the last non-zero disturbance word may appear in position 55, because we need eight steps 56,...,63 to correct it. thus, the last 8 words of the expanded difference $E \cdot \Delta_M$ have to be zero. Since E is defined by (5), this condition can be written as (6).

Now, let us consider the following linear combination of $\Delta$ and its delayed versions

$C = \Delta - 4\,\text{Delay}^1 64(\Delta) + 2\,\text{Delay}^2 64(\Delta) + 2\,\text{Delay}^3 64(\Delta) + 4\,\text{Delay}^4 64(\Delta) + 2\text{Delay}^5 64(\Delta) + \text{Delay}^6 64(\Delta) - \text{Delay}^8 64(\Delta)$ **(8)**

It is easy to see that each disturbance word $\Delta_I$ in C is corrected by its appropriate multiplicities appearing in the next eight positions and coming from the delayed vectors.

Since the message expansion is linear, C is the result of the expansion if and only if all the delayed and truncated vectors Delay$^b$64($\Delta$), $0 \leq b \leq 8$ are results of the expansion process.

LEMMA 3.1

Assures that it is true if Delay$^8$($\Delta$) = $[0,0,0,0,0,0,0,0,\Delta_0,...,\Delta_{63}]^T$ is the result of the (extended, N= 68) expansion process. We can achieve this by taking the first 16 words and expanding them forward according to Equation (1), but also by taking any 16 consecutive words and expanding partly forward and partly backward. In our case we select elements 8–23 for the expansion. If we index elements of Delay$^8$($\Delta$) starting from -8 and split the vector into two parts: one having negative and the other one having non-negative indices, we can express this requirement equivalently by the following two conditions:

$[\cdot,\cdot,\cdot,\cdot,\cdot,\cdot,\cdot,\cdot,0,0,0,0,0,0,0,0]^T$ = $A^{-1} \Delta_M$ and $\Delta = E \cdot \Delta_M$. Only the first condition, namely $A^{-1}\Delta_M$ has to end with 8 zeros, has to be satisfied, since $\Delta$ is already the result of an expansion. This condition can be written simply in the form of Equation (7) what completes the proof. As long as Equations (6) and (7) are satisfied, $\Delta$ is a valid disturbance pattern and C is a complete differential characteristic corresponding to it.

After obtaining explicit forms of the matrices $A^3$ and $A^{-1}$ (this is possible since A is a bijection) we solve the system of equations given by (6) and (7) over $Z_{2^{32}}$ and get the following result

$$\Delta_M = [\text{0x10000000, 0xa0000000, 0xc0000000, 0xa0000000,}$$
$$\text{0xe0000000, 0x20000000, 0x40000000, 0x40000000,}$$
$$\text{0x80000000, 0xd0000000, 0x10000000, 0x60000000,}$$
$$\text{0x50000000, 0x40000000, 0x70000000, 0x30000000}]^T$$
(9)

This shows that the solution space is just one-dimensional. Any multiple of $\Delta_M$ is also a solution, but since all components of the vector (9) have only up to four most significant binary digits different from zero (so they are all of the form $a_i \cdot 2^{28} \pmod{2^{32}}$ where $a_i \in \{0,...,15\}, 0 \leq i < 16$), there are only 16 distinct disturbance patterns. using any of them results in collision for ADD-linearised SHA-256

IV. INCORPORATING BOOLEAN FUNCTIONS

Now let us consider a variant of SHA-256 still without S-Boxes, but with both Boolean functions M$aj$ and Ch in place. If we multiply the basic pattern (9) by 8 (so shift it left by 3 bit positions), we get a disturbance pattern $\Delta_* = E$ ($8\Delta_M$) that has non-zero bits at the most significant bits only. The most significant bits of $\Delta_*$ are as follows

10000000011010111011100110100110  (10)
0000011100101111 1011100000000000

$\Delta_*$ is a disturbance pattern that not only follows the message expansion but also allows us to treat it as a binary pattern with a relatively low weight of 27.

We can approximate both Boolean functions with probability at least ½ assuming that the function produces an output difference each time the input difference is non-zero. This approximation is shown in Table 2.

Table 2: Probabilities of non-zero output differences for the Boolean functions Ch and M aj

| input difference $(\delta_x, \delta_y, \delta_z)$ | $Ch$ function conditions | prob | $Maj$ function conditions | prob |
|---|---|---|---|---|
| (1,0,0) | $y + z = 1$ | 1/2 | $y + z = 1$ | 1/2 |
| (0,1,0) | $x = 1$ | 1/2 | $x + z = 1$ | 1/2 |
| (0,0,1) | $x = 0$ | 1/2 | $x + y = 1$ | 1/2 |
| (1,1,0) | $x + y + z = 0$ | 1/2 | $x + y = 0$ | 1/2 |
| (1,0,1) | $x + y = 0$ | 1/2 | $x + z = 0$ | 1/2 |
| (0,1,1) | – | 1 | $y + z = 0$ | 1/2 |
| (1,1,1) | $y + z = 0$ | 1/2 | – | 1 |

we use this approximation and trace how a single bit disturbance $\Delta_{*i}$ introduced in step i propagates through the next 8 steps, we get the following sequence of corrections: $\{0,0,\Delta_i,\Delta_i,0,0,0,\Delta_i\}$,  (11)

Which we need in steps i +1,...,i+8 in order to cancel the initial disturbance $\Delta$. The whole process is very similar to the one used to obtain the sequence of corrections (as given in (4)).

A complete differential is obtained in the same way as in the previous case, by adding delayed disturbance patterns multiplied by corresponding coefficients of Equation (11), i.e. $\{0,0,1,1,0,0,0,1\}$

This time however, correction process is probabilistic as each active Boolean function almost always (except for input differences (0, 1, 1) for Ch and (1, 1, 1) for M aj) introduces a factor of 1/2. A detailed analysis of these probabilities is presented in Table 3.

After multiplication of all factors, we obtain a probability for a successful correction equal to $2^{-84}$. Further optimisation are also possible as we can choose messages in such a way that conditions for successful correction will be always satisfied for the first 16 steps, what could increase the probability to around $2^{-64}$. This shows that the use of substitution boxes $\sigma_0$, $\sigma_1$ and $\Sigma_0$, $\Sigma_1$ is essential for the security of SHA-256 and also demonstrates that mixing only modular additions with Boolean functions is not enough for constructing a secure hash function.

| s | Maj | Ch | e | s | Maj | Ch | e | s | Maj | Ch | e | s | Maj | Ch | e |
|---|---|---|---|---|---|---|---|---|---|---|---|---|---|---|---|
| 0 | 000 | 000 | 0 | 16 | 110 | 010 | 2 | 32 | 011 | 100 | 2 | 48 | 111 | 110 | 1 |
| 1 | 100 | 100 | 2 | 17 | 111 | 101 | 1 | 33 | 001 | 010 | 2 | 49 | 111 | 011 | 0 |
| 2 | 010 | 010 | 2 | 18 | 011 | 010 | 2 | 34 | 000 | 001 | 1 | 50 | 011 | 101 | 2 |
| 3 | 001 | 101 | 2 | 19 | 101 | 001 | 2 | 35 | 000 | 100 | 1 | 51 | 101 | 010 | 2 |
| 4 | 000 | 110 | 1 | 20 | 110 | 100 | 2 | 36 | 000 | 010 | 1 | 52 | 110 | 101 | 2 |
| 5 | 000 | 111 | 1 | 21 | 111 | 110 | 1 | 37 | 000 | 001 | 1 | 53 | 111 | 110 | 1 |
| 6 | 000 | 011 | 0 | 22 | 011 | 011 | 1 | 38 | 100 | 100 | 2 | 54 | 011 | 011 | 1 |
| 7 | 000 | 001 | 1 | 23 | 001 | 101 | 2 | 39 | 110 | 110 | 2 | 55 | 001 | 101 | 2 |
| 8 | 000 | 000 | 0 | 24 | 100 | 110 | 2 | 40 | 111 | 011 | 0 | 56 | 000 | 010 | 1 |
| 9 | 000 | 000 | 0 | 25 | 110 | 011 | 1 | 41 | 011 | 001 | 2 | 57 | 000 | 101 | 1 |
| 10 | 100 | 000 | 1 | 26 | 011 | 101 | 2 | 42 | 001 | 100 | 2 | 58 | 000 | 010 | 1 |
| 11 | 110 | 110 | 2 | 27 | 101 | 110 | 2 | 43 | 100 | 110 | 2 | 59 | 000 | 001 | 1 |
| 12 | 011 | 111 | 2 | 28 | 010 | 011 | 1 | 44 | 010 | 111 | 2 | 60 | 000 | 000 | 0 |
| 13 | 101 | 111 | 2 | 29 | 001 | 111 | 2 | 45 | 101 | 011 | 1 | 61 | 000 | 000 | 0 |
| 14 | 010 | 011 | 1 | 30 | 100 | 000 | 1 | 46 | 110 | 001 | 2 | 62 | 000 | 000 | 0 |
| 15 | 101 | 101 | 2 | 31 | 110 | 000 | 1 | 47 | 111 | 100 | 1 | 63 | 000 | 000 | 0 |

## V. THE ROLE OF S-BOXES

The substitution boxes $\Sigma_0$ and $\Sigma_1$ constitute the essential part of the hash function and fulfil two tasks: they add bit diffusion and destroy the ADD-linearity of the function.

There are modular differentials for $\Sigma_0$ and $\Sigma_1$ that hold for one bit input difference e with probability $2-3$ (necessary for S-boxes used in steps $i+$ , $i+5$ and with probability around $2-10$ for input difference equal to $\Sigma_0(e)$ (used for $\Sigma_1$ in step $i+2$). using the approach of modular differences it is possible to obtain a corrective pattern for the complete round structure with probability around $2-42$. A better result of $2-39$ was obtained by Hawkes *etal* [HPR04] by explicit computation of modular differences for $\Sigma_0$ and $\Sigma_1$, rather than approximating them with a constant differential.

The S-boxes $\sigma_0$ and $\sigma_1$ play a similar role: they provide non linearity and better diffusion for the message expansion.

These two properties of the message expansion constitute the foundation of the security of the full SHA-256, as in order to apply corrective patterns in a straightforward way, one would need at least 37 expanded words equal to zero (since at most three corrective patterns can be applied). Although this seems to be unlikely, further research is needed in this direction. In the rest of this section, we concentrate on the message expansion and list some interesting properties of it:

- $\sigma_0$ and $\sigma_1$ have both the property to increase the hamming weight of low-weight inputs. This increase is upper bounded by a factor of 3. The average increase if Hamming weight for low-weight inputs is even higher if three rotations are used instead of two rotations and one bit-shift. However, a reason for this bit-shift is given by the next observation.

- In contrast to all other members of the MD4-family including SHA-1, rotating expanded message words to get new expanded message words is not possible anymore (even in the XOR- linearised case). This is due to the bit-shift being used in $\sigma_0$ and $\sigma_1$.

## VI. FINDING LOW-WEIGHT CODE WORDS IN THE CODE DESCRIBING THE XOR- LINEARISED SHA-256 MESSAGE EXPANSION

In the first attempt to get an idea about the effect of all the changes between the SHA-1 message expansion and the SHA-256 message expansion, we consider single bit differences. Table 4 illustrates this comparison. We consider variants reduced to 40 steps as well as full variants (80 steps for SHA-1 variants and 64 steps for SHA-256 variants).

By the modified SHA-1 message expansion we refer to a variant where every XOR is replaced by an addition modulo $2^{32}$. By the modified SHA-256 message expansion, we refer to a variant where every addition is replaced by an XOR. We observe that both the introduction of modular additions and the replacement of a single bit-shift by a structure bit difference in various message expansions. In variants using modular addition, we used the all-zero vector as a starting point. using $\sigma_0$ and $\sigma_1$ heavily increases the number of affected bits in the expanded message.

|  | orig. SHA-1 | mod. SHA-1 | mod. SHA-256 | orig. SHA-256 |
|---|---|---|---|---|
| min (40 steps) | 18 | 18 | 110 | 137 |
| max (40 steps) | 30 | 41 | 297 | 307 |
| min (full) | 107 | 247 | 467 | 507 |
| max (full) | 174 | 354 | 694 | 709 |

*Table 4: Comparison of the number of affected bits for a single*

When talking about the SHA-1 message expansion, it was already observed in the works [MP05, RO05] that weights much smaller than 107 (as given in Table 4) can be found. The minimum weight found for the message expansion of SHA-1 is 44. A more recent treatment of low-weight disturbance patterns in SHA-1 can be found in [JP05].

Due to the nonlinear behaviour of the modular addition, no linear code can describe the SHA-256 message expansion. However, if the modular addition is replaced by XOR, a linear code over $Z_2$ can be constructed. If we consider SHA-256 with N steps, this code can be represented by a 512 × 32N generator matrix.G.

Due to the XOR-linearisation, every possible difference of two expanded words is also a valid word in this code. Therefore, probabilistic algorithms from coding theory [Leo88, Ste89, CC98] can be used to find low-weight differences for the XOR-linearised SHA-256 message expansion. some results of this codeword search are depicted in Figure 2. All minimum weights found for variants of the message expansion up to the full 64 steps are shown in the figure. Until the 42-step variant, our algorithms found reasonable low weights. This is depicted by the solid line. Considering the 40-step variant, the weight of 26 is low compared to a minimal weight of 110 for single-bit differences given in Table 4. The 40-step expanded message is given in Table 5.

For variants with more than 42 steps, the running time of our algorithms is currently too high to return reasonable low weights. The sudden jump after step 42 is not an intrinsic property of the SHA-256 message expansion, but rather the result of the limited running time of our algorithms.

To show that there indeed are low-weight words for N > 42, we proceed as follows. After obtaining a low-weight word for 42 steps we use the expansion process to extend it to the full length word. weights obtained in this way are depicted by the dashed line.

A 42-step word of weight 35 is used there as a starting point. Expanding it to 64 steps gives us a weight of 356. This is considerable lower than 467, which is the minimal weight given for a single bit difference in Table 4.

However, there is room for improvements.

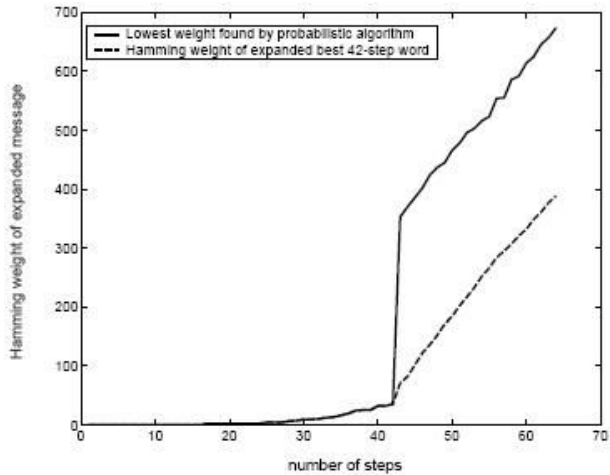

Figure 2: Hamming weights of low-weight words found for step-reduced variants of the XOR- linearised SHA-256 message expansion.

| 00000001 | 00040088 | 00000000 | 00000000 |
| 00000000 | 00000001 | 00000000 | 00000000 |
| 00000000 | 15522028 | 00000000 | 00000000 |
| 00000000 | 000A0400 | 00000000 | 00000000 |
| 00000000 | 00000000 | 00000000 | 00000000 |
| 00000000 | 00000000 | 00004050 | 00000000 |
| 00000000 | 00000000 | 00000000 | 00000000 |
| 00000000 | 00040088 | 00000001 | 00000000 |
| 00000000 | 00000001 | 00000000 | 00000000 |
| 00000001 | 00000000 | 00000000 | 00000000 |

In contrast to the words found for the SHA-1 message expansion, there are no zero bands [RO05] any more. Note that the given expanded message is not necessarily a valid difference in case of the real message expansion since we approximate the modular addition by the bitwise XOR operation. Also note that the given vector cannot directly be used as a collision-producing disturbance pattern as described by Chabaud and Joux in their original attack on SHA-0 [CJ98]. The reason is that there are *truncated local collisions* [CJ98] generated by non-zero words in the backward expansion. These local collisions start before step 0 and would cause additional difficulties for constructing a collision-producing differential characteristic. However, we expect to find input words for reduced variants of the message expansion that can be used to build a collision-producing difference.

A number of conditions on chaining variables need to be satisfied in order to ensure that the concatenation of local collisions (which hold with a probability between $2^{-39}$ and $2^{-42}$) results in a collision of the output of the compression function. If we do not assume any pre-fulfilled conditions, the maximal weight we allow for a perturbation pattern is 3 (since $2^{-39 \cdot 4} < 2^{-128}$). Considering the weights in Figure 2, this would mean a maximum of 24 steps expansion and presented expanded messages with low Hamming weights for the XOR- linearised message expansion of SHA-256. The general ideas of all these results apply also to other members of the SHA-2 family.

## VII CONCLUSIONS

In this paper we presented methods for finding collisions for two simplified variants of SHA-256, one fully linearised with respect to the modular addition and the other one with all the S-Boxes replaced by the identity function. These results show that the presence of S-Boxes is essential for the security of SHA-256. We studied properties of the message

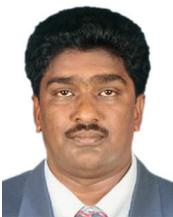

*A.Arul Lawrence selvakumar received the M.E., M.S.,degrees in Computer Science & Engineering from Bharathidasan University, INDIA and pursuing his Ph.D., in cryptography. He is currently working as a Professor in the Dept / CSE at Royal College of Engineering & Technology,Thrissur,Kerala ,India.He is member of various professional societies like CSTA, IAENG, and IACSIT. He has published more than 26 papers in international journals and conference and also published 2 books in network security. He has attended more than 9 workshops, symposia, seminars and short-term training programmes. His research interest includes network security, cryptography and client server computing.*